# Broadband Free Space Impedance in Co₂Z Hexaferrites by Substitution of Quadrivalent Heavy Transition Metal Ions for Miniaturized RF Devices

Piotr Kulik[1*], Gavin Winter[1], Alexander Sokolov[1], Katherine Murphy[1], Chengju Yu[1], Kun Qian[1], Ogheneyunume Fitchorova[2], and Vincent Harris[1]

[1]Center for Microwave Magnetic Materials and Integrated Circuits (CM³IC), Department of Electrical and Computer Engineering, Northeastern University, Boston, MA 02115 USA
[2]KRI at Northeastern University, LLC, Burlington, MA 01803 USA

*Contact author: kulik.p@husky.neu.edu

*Abstract*- Polycrystalline samples of Z-type hexaferrites, having nominal compositions $Ba_3Co_{2+x}Fe_{24-2x}M_xO_{41}$ where M = $Ir^{4+}$, $Hf^{4+}$, or $Mo^{4+}$ and x=0 and 0.05, were processed via ceramic processing protocols in pursuit of low magnetic and dielectric losses as well as equivalent permittivity and permeability. Fine process control was conducted to ensure optimal magnetic properties. Organic dispersants (i.e., isobutylene and maleic anhydride) were employed to achieve maximum densities. Crystallographic structure, characterized by X-ray diffraction, revealed that doping with $Ir^{4+}$, $Hf^{4+}$, or $Mo^{4+}$ did not adversely affect the crystal structure and phase purity of the Z-type hexaferrite. The measured microwave and magnetic properties show that the resonant frequency shifts depending on the specific dopant allowing for tunability of the operational frequency and bandwidth. The frequency bandwidth in which permittivity and permeability are very near equal (i.e., ~400 MHz for $Mo^{4+}$(x), where x=0.05 doping) is shown to occur at frequencies between 0.2 and 1.0 GHz depending on dopant type. These results give rise to low loss, i.e., $\tan \delta_\varepsilon / \varepsilon' = 0.0006$ and $\tan \delta_\mu / \mu' = 0.038$ at 650 MHz, with considerable size reduction of an order of magnitude, while maintaining the characteristic impedance of free space (i.e., 377±5Ω). These results allow for miniaturization and optimized band-pass performance of magnetodielectric materials for communication devices such as antenna and radomes that can be engineered to operate over desired frequency ranges using cost effective and volumetric processing methodologies.

*Keywords: Z-type hexaferrites, radio frequency, impedance engineering, equivalent permittivity and permeability, magnetic and dielectric loss*

The demand for miniaturized and high performing broadband communication components and systems in the MHz-GHz frequency bands is steadily growing due to a severely crowded and rapidly evolving commercial and military spectral environment.[1]

As a result, advanced magnetodielectric materials are necessary to enable high permeability and permittivity with low magnetic losses and broadband performance[2]. Using magnetodielectric materials, with properties such as equivalent $\varepsilon'$ and $\mu'$, allows for the design of miniaturized components for modern communication systems such as radomes, antennas[3] and electromagnetic band-gap (EBG) substrates.[4] Additionally, $\varepsilon' = \mu'$ enables impedance matching to free space, an important parameter for antennas to operate without impedance matching circuitry as well as allowing for efficient power transfer and band-pass operation.[5,6]

In this work, a series of Z-type barium hexaferrites were prepared in which dopants of quadrivalent heavy transition metal ions, i.e., $Mo^{4+}$, $Ir^{4+}$, and $Hf^{4+}$, were employed to modify microwave properties. Through systematic processing and measurements, it was revealed that each dopant affected the hexaferrites' properties differently by both shifting of the resonant frequency and altering magnetic properties without strongly affecting the hexaferrite crystallographic structure[7]. Additionally, $\varepsilon' = \mu'$ was realized without any additives such as $Bi_2O_3$, which had been previously used for tuning permeability and permittivity in hexaferrite systems.[8] Hence, this allowed for a cost effective approach and ease of manufacturing of hexaferrite materials for highly multifunctional and miniaturized radio frequency devices and systems.

Polycrystalline $Co_2Z$ hexaferrites[9,10], having a nominal composition of $Ba_3Co_{2+x}Fe_{24-2x}M_xO_{41}$ where M = $Ir^{4+}$, $Hf^{4+}$, or $Mo^{4+}$ and x=0 and 0.05 were prepared by ceramic processing protocols. $BaCo_3$, $IrO_2$, $HfO_2$, $MoO_2$, $Co_3O_4$, and $Fe_2O_3$ of high purity (i.e., ≥99.95%) (Sigma-Aldrich) were mixed and calcined in flowing oxygen gas at a rate of 1 standard cubic feet per hour (SCFH) for 5-8 hours at 900-1000 °C. The calcined mixtures were then ball milled for 18-24 hours to a particle size ranging from 1-2 μm. Achieving a particle size of 1–2 μm provides for the stabilization of a single-magnetic domain within the particle that allows for low magnetic coercivity and hysteretic losses.[11,12] The grain size was measured by scanning electron microscopy (FEI Scios Dual-Beam FESEM). Furthermore, isobutylene and maleic anhydride (Kuraray ISOBAM 104) with 70 to 80% hexaferrite solid loading were combined with deionized water to form a slurry that was mixed at 2500 rpm for several minutes. The addition of ISOBAM allowed for the samples to reach a maximum density close to the theoretical value of 5.37 g cm$^{-3}$.[13] Samples were then pressed to form toroids having an inner diameter of 3 mm and an outer diameter of 7 mm. This sample size was chosen for characterization within transmission air-line microwave measurements (Maury Microwave No.2650CK). The toroids were then additionally compacted at a pressure of 3.45x10$^7$ Pa under cold isostatic conditions to maximize density prior to sintering. Sintering was performed for 4-6 hours at 1000-1250 °C in an oxygen gas environment; conditions that have been demonstrated to reduce magnetic and dielectric losses.[14]

Crystallographic structure was determined using a θ–2θ powder X-ray diffractometer (Rigaku, Ultima III) at room temperature employing Cu k$\alpha$ radiation (λ=1.5406 Å). The complex permittivity and permeability spectra were measured over a frequency range from 0.2 to 4 GHz using an Agilent E864A 45MHz–50GHz PNA series network analyzer and a 7 mm HP 85050C precision airline. In this Letter, we present and discuss measured values of complex permeability and permittivity, impedance, and losses for these novel materials.

In order to optimize RF properties, i.e., magnetization, low losses[15], and high permittivity and permeability, the grain size of the ferrite samples together with their density, were optimized by varying process conditions such as milling and sintering protocols. Scanning electron microscopy was employed to measure sample microstructure and particle morphology. It can be seen from Fig. 1(a) that the grain size after 8 hrs. of ball milling, prior to sintering, was ~ 5 μm, however there exists some grains that are larger. In order to minimize this bimodal grain size distribution, an additional 24-hour ball milling step was employed. Figure 1(b) illustrates the final grain size that resulted from the 24-hour ball milling step prior to sintering. Control of grain size distribution allows for the optimization of magnetic properties for the substituted Co$_2$Z hexaferrites. This is seen in Fig. 1(c), which shows the hexaferrite particle size to range from ~ 3-20 μm after sintering. The wide range in particle size reflects nucleation (i.e., early) and mature stages of grain growth and ripening.

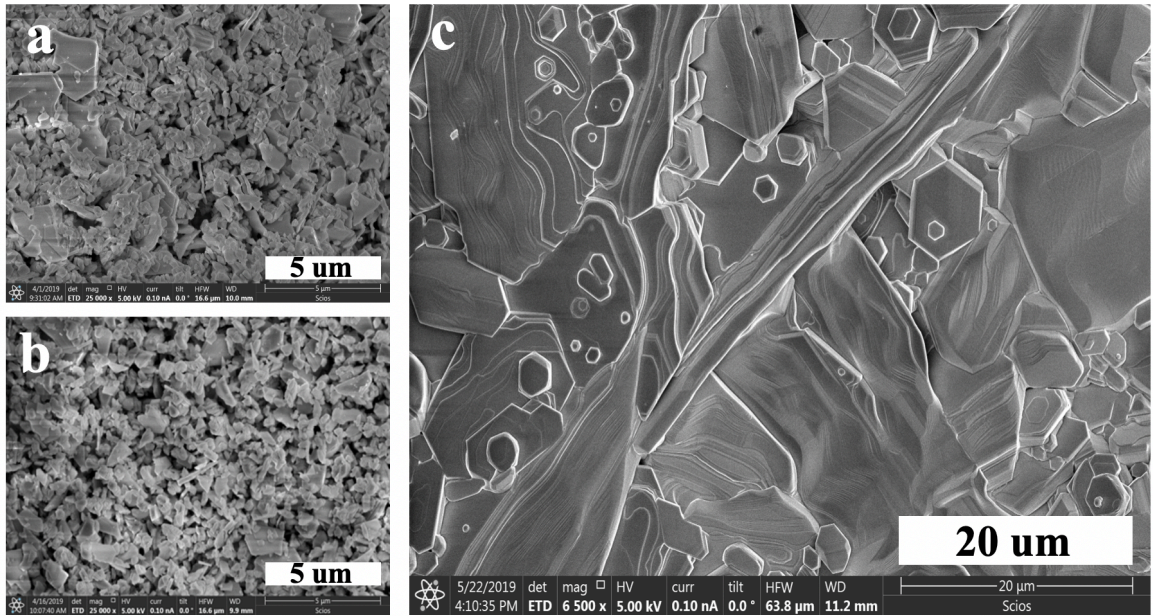

**Figure 1.** Scanning electron microscopy images of (a) pre-sintered powder after 8 h of ball milling and (b) of a toroid sample created after a 24-hour ball milling cycle, and (c) a final toroid sample with large hexaferrite crystals ranging from a diameter of approximately 3-20 μm (see text for discussion).

X-ray diffraction (XRD) was performed in order to characterize the crystal structure of all hexaferrite samples of the present study. Experimental XRD patterns, with similar data presented as diffraction lines of intensity and 2θ placement extracted from the JCPDS data file corresponding to Z-type and Y-type hexaferrite phases are shown in Fig. 2. In this figure, all diffraction features are indexed to either Z-type or Y-type hexaferrite crystallographic phases. Z-type hexaferrite materials prepared by solid state reaction often contain a secondary phase of W- and/or Y-type phases[10]. However, in Fig. 2 nearly all reflections of the Y-type and Z-type phases overlap making it difficult to ascertain the presence and volume fraction of a Y-type hexaferrite phase. However, the (119) peak having a relative intensity in the Y-type XRD spectrum of 74% (JCPDS PDF #44-0206) appears at 2θ~36°. The presence of this peak in all samples does not rise above $\simeq 1-2$ % relative intensity, a value consistent with a Z-type phase, indicating that the samples under study are pure Z-type hexaferrites within the detection sensitivity of the diffractometer.

Generally, in magnetoceramics of appropriate nominal compositions heat-treatments at temperatures ranging between 1200-1300 °C have been shown to result in pure Z-type phase.[13] Hence, similar heat treatment conditions have been employed here. It is seen in Fig. 2, that the parent composition, which does not include dopants, and the compositions that have been modified by low levels of $Mo^{4+}$, $Ir^{4+}$, and $Hf^{4+}$ dopants do not demonstrate measurable changes to the crystallographic phase purity[16]. Intensities of the Z-type peak change slightly from the parent composition (i.e., x=0) to those containing dopants; however, the Y-type secondary phase diffraction features remains below the detection limits and therefore one may conclude that the introduction of dopants do not significantly change the crystal structure from that of a single Z-type phase.

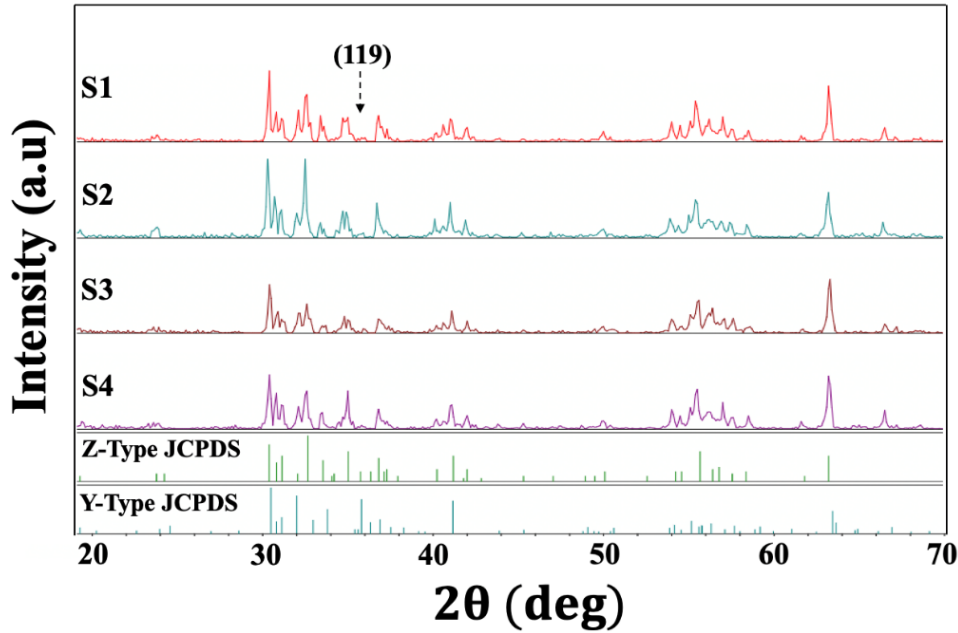

**Figure 2. X-ray diffraction patterns of ferrite samples (S1: $Ba_3Co_2Fe_{24}O_{41}$, S2: $Ba_3Co_{2.05}Hf_{0.05}Fe_{23.90}O_{41}$, S3: $Ba_3Co_{2.05}Mo_{0.05}Fe_{23.90}O_{41}$, and S4: $Ba_3Co_{2.05}Ir_{0.05}Fe_{23.90}O_{41}$). Additionally, standard diffraction information extracted from relevant JCPDS data cards corresponding to Z-type and Y-type hexaferrite features are included. Data were collected at room temperature in a θ–2θ geometry using a Cu k$\alpha$ radiation source.**

In order to investigate the affect upon microwave properties of substitutions[17] on the Z-type hexaferrites, the complex permittivity and permeability spectra were measured over a frequency range from 0.2 to 4.0 GHz using an Agilent E864A 45MHz–50GHz PNA series network analyzer and a 7 mm HP 85050C precision airline.

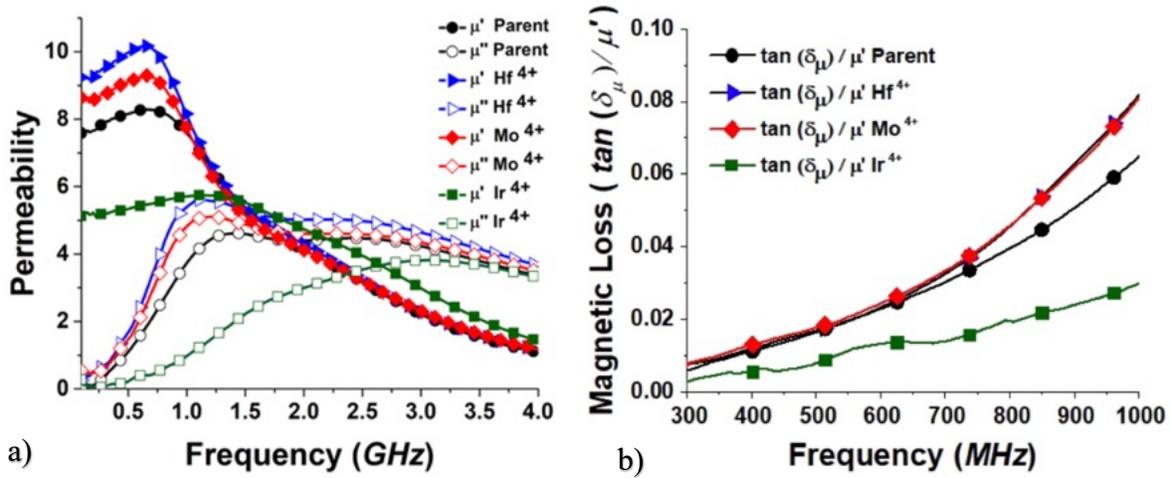

**Figure 3.** (a) Complex permeability ($\mu'$ and $\mu''$) and (b) magnetic loss tangents ($\tan(\delta_\mu)/\mu'$) of the parent and doped $Ba_3Co_{2+x}Fe_{24-2x}M_xO_{41}$ samples. One sees that the operational band is near 0.20 to 0.80 GHz.

As displayed in Fig. 3 (a), the permeability changes substantially with the addition of various dopants. From Fig. 3 (a), it can be seen that upon substitution of $Ir^{4+}$ ions to the parent composition the maximum value of µ' shifts to higher frequencies resulting in a lower permeability of ~6 at 1.25 GHz. However, when $Hf^{4+}$ is added to the parent composition the maximum value of µ' shifts to lower frequencies resulting in a higher permeability ~10 at 0.70 GHz. Results of adding $Mo^{4+}$ ions have the maximum value of µ' shifting in between that of the other phases with dopants of $Ir^{4+}$ and $Hf^{4+}$ with permeability and resonant frequencies falling in the middle of the spectra corresponding to the different dopants. Additionally, as permeability decreases the magnetic loss tangent decreases as seen in the case of $Ir^{4+}$ shown in Fig. 3 (b). This behavior corresponds to the quadrivalent heavy transition metal cations on both frequency and permeability allows the customization of the ferrite for specific applications with low loss requirements. The effect of the dopants can be attributed in part to the ionic radii of the dopants and the corresponding distortion to the crystal lattice and its influence upon the magnetocrystalline anisotropy energy[18] and fields. Changes to magnetic anisotropy energy play a major role in the gyromagnetic properties of ferrites in determining conditions and performance of operation.

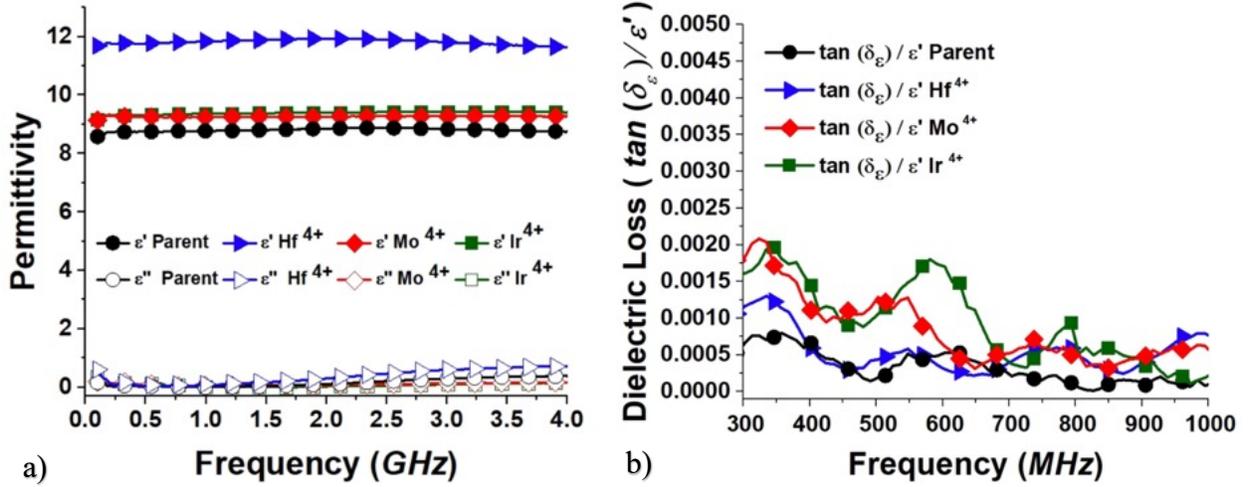

**Figure 4.** (a) Complex permittivity ($\varepsilon'$ and $\varepsilon''$) and (b) dielectric loss tangents (tan ($\delta_\varepsilon$)/$\varepsilon'$) of the parent and doped $Ba_3Co_{2+x}Fe_{24-2x}M_xO_{41}$ samples.

Additionally, the dopant impact on complex permittivity values and dielectric loss tangents is examined. From Fig. 4 (a) it can be seen that the parent, $Mo^{4+}$, and $Ir^{4+}$ samples have $\varepsilon'$ values of about 9, however the $Hf^{4+}$ shows a $\varepsilon'$ value of nearly 12. This supports that $Hf^{4+}$ can serve better as a dopant in terms of device size reduction if the permeability values improve, which is discussed later in this Letter. Analyzing complex permittivity also allows for the determination of dielectric loss tangents, which are an important parameter to consider when designing RF devices. From Fig. 4 (b), the dielectric loss tangent is seen to be below 0.0020 for all dopants, with the parent sample showing the lowest values. The variation in dielectric constant over this frequency range is attributed to variations in density between each sample. The measured densities of each sample do not exceed 4% in variation from the theoretical value of 5.37 g cm$^{-3}$, e.g., the densities of the parent, $Hf^{4+}$, $Mo^{4+}$, and $Ir^{4+}$ are 5.29 g cm$^{-3}$, 5.28 g cm$^{-3}$, 5.13 g cm$^{-3}$, and 5.17 g cm$^{-3}$, respectively.

To further investigate the properties and application value of these hexaferrites, especially in applications such as antennas and radomes, it is critical to understand the role of impedance and form factor[19]. In order to achieve free space impedance matching it is important to realize equivalent values of permeability and permittivity as shown for the $Mo^{4+}$ doped sample in Fig. 5(a). The impedance (Z) and size reduction factor (SRF) equations are expressed as Eqns. (1) and (2), respectively:

$$Z = \sqrt{\frac{\mu_o \mu_r}{\varepsilon_o \varepsilon_r}} \quad (1) \qquad\qquad SRF = \sqrt{\varepsilon_r' \mu_r'} \quad (2)$$

where $\mu_o$, $\varepsilon_o$, $\mu_r$, and $\varepsilon_r$ are the permittivity and permeability in vacuum and of the hexaferrite, respectively. By applying Eqns. (1) and (2), the impedance of each doped hexaferrite sample was determined and presented in Fig. 5(b). The $Mo^{4+}$ doped sample maintains the largest free space impedance bandwidth of samples studied here, ~ 400 MHz, at a center frequency of ~ 650 MHz.

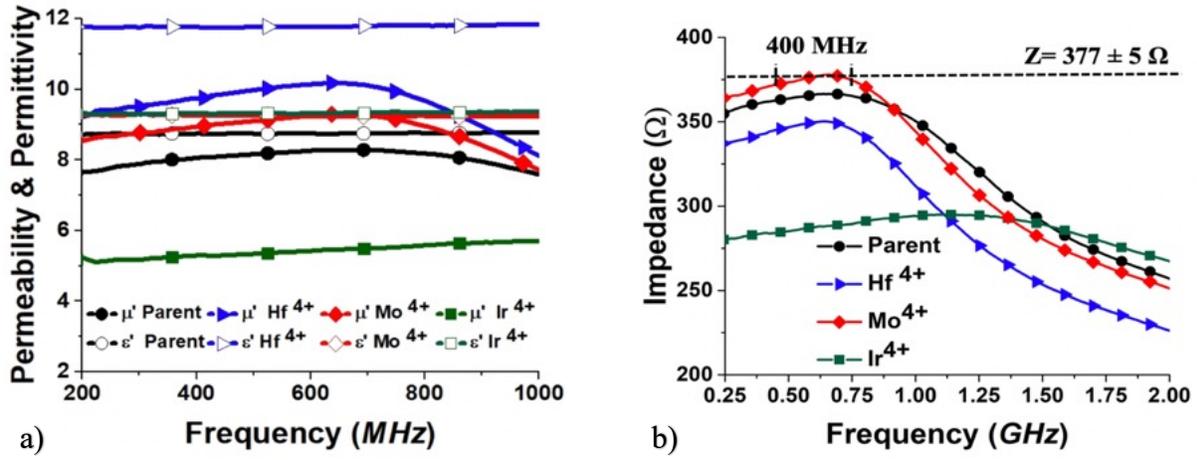

**Figure 5.** (a) Permeability and permittivity ($\mu'$ and $\varepsilon'$) as a function of frequency for the parent and doped $Ba_3Co_{2+x}Fe_{24-2x}M_xO_{41}$ samples. (b) Impedance (Z). The impedance of free space, 377 Ω, is achieved when permeability and permittivity are made equal as is demonstrated in panel (b) for $Mo^{4+}$ doping (x=0.05) where Z=377±5 Ω (see solid red curve with diamond symbols).

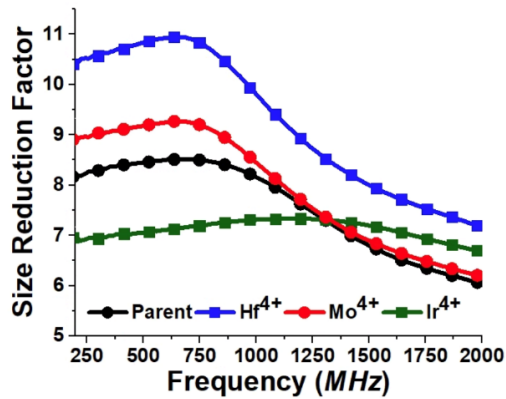

**Figure 6.** Size reduction factor as a function of frequency for the doped $Ba_3Co_{2+x}Fe_{24-2x}M_xO_{41}$ samples.

Obtaining higher values of permeability and permittivity allows for larger size reduction as derived from Eqn. (2). Measured values are displayed in Fig. 6. A size reduction factor greater than 9 is achieved in the impedance matching band of the Mo$^{4+}$ doped sample. This occurs at a center frequency of ~650 MHz[20] with a minimum in dielectric and magnetic loss of tan $\delta_\varepsilon/\varepsilon' = 0.0006$ and tan $\delta_\mu/\mu' = 0.038$, respectively. This particular material is a promising candidate for miniaturized antenna designs intended for operation at UHF frequencies.

In summary, polycrystalline Co$_2$Z hexaferrites, having a nominal composition of Ba$_3$Co$_{2+x}$Fe$_{24-2x}$M$_x$O$_{41}$ where M = Ir$^{4+}$, Hf$^{4+}$, or Mo$^{4+}$ and x=0 and 0.05 were prepared by solid state ceramic processing protocols. SEM images reveal that the particle sizes after sintering were around 3-20 μm, which are consistent with single magnetic domain particles. XRD analyses show that as the dopant type and amount were changed there appeared a small amount ($\simeq$1-2%) of secondary Y-type phase. However, the magnetic properties and frequency of operation were markedly different. These results show that the impedance of Co$_2$Z ferrites can be tuned using quadrivalent heavy transition metal substitution to that of free space, i.e., 377±5 Ω, at frequencies in the UHF band, while maintaining low losses with considerable size reduction. As a result, antenna, EBG substrates, and radomes, among other applications, can be developed for different applications in the UHF where unique RF behavior and device miniaturization are desirable.

[20] Cassella, C., Assylbekova, M., Zhu, W. Z., Chen, G., Kulik, P., Michetti, G, & Rinaldi, M. (2018, June). 750 MHz Zero-Power MEMS-Based Wake-Up Receiver with-60 dBm Sensitivity. *In Hilton Head Workshop (pp. 3-7).*